\makeatletter
\newcommand{\dontusepackage}[2][]{%
  \@namedef{ver@#2.sty}{9999/12/31}%
  \@namedef{opt@#2.sty}{#1}}
\makeatother
\dontusepackage{subfigure}

\documentclass[]{article}

\usepackage{lmodern}
\usepackage{amssymb,amsmath}
\usepackage{ifxetex,ifluatex}
\usepackage[usenames,dvipsnames]{color}
\usepackage{fixltx2e} 
\ifnum 0\ifxetex 1\fi\ifluatex 1\fi=0 
  \usepackage[T1]{fontenc}
  \usepackage[utf8]{inputenc}
\else 
  \ifxetex
    \usepackage{mathspec}
    \usepackage{xltxtra,xunicode}
  \else
    \usepackage{fontspec}
  \fi
  \defaultfontfeatures{Mapping=tex-text,Scale=MatchLowercase}
  
\fi
\IfFileExists{upquote.sty}{\usepackage{upquote}}{}
\IfFileExists{microtype.sty}{%
\usepackage{microtype}
\UseMicrotypeSet[protrusion]{basicmath} 
}{}
\usepackage[margin=1.0in,bottom=1.5in]{geometry}
\usepackage[]{natbib}
\usepackage{listings}
\usepackage{graphicx}
\makeatletter
\def\maxwidth{\ifdim\Gin@nat@width>\linewidth\linewidth\else\Gin@nat@width\fi}
\def\maxheight{\ifdim\Gin@nat@height>\textheight\textheight\else\Gin@nat@height\fi}
\makeatother
\setkeys{Gin}{width=\maxwidth,height=\maxheight,keepaspectratio}
\usepackage{caption}
\usepackage{float}
\usepackage{subfig}
\ifxetex
  \usepackage[setpagesize=false, 
              unicode=false, 
              xetex]{hyperref}
\else
  \usepackage[unicode=true]{hyperref}
\fi
\hypersetup{breaklinks=true,
            bookmarks=true,
            pdfauthor={},
            pdftitle={Learned coupled inversion for carbon sequestration monitoring and forecasting with Fourier neural operators},
            colorlinks=true,
            citecolor=black,
            urlcolor=blue,
            linkcolor=black,
            pdfborder={0 0 0}}
\title{Learned coupled inversion for carbon sequestration monitoring and
forecasting with Fourier neural operators}
\author{Ziyi Yin, Ali Siahkoohi, Mathias Louboutin, Felix J. Herrmann\\Georgia
Institute of Technology\\}
\date{}

\begin{document}
\maketitle

\vspace*{-0.3cm}

\section{Summary}\label{summary}

\vspace*{-0.3cm}

Seismic monitoring of carbon storage sequestration is a challenging
problem involving both fluid-flow physics and wave physics.
Additionally, monitoring usually requires the solvers for these physics
to be coupled and differentiable to effectively invert for the
subsurface properties of interest. To drastically reduce the
computational cost, we introduce a learned coupled inversion framework
based on the wave modeling operator, rock property conversion and a
proxy fluid-flow simulator. We show that we can accurately use a Fourier
neural operator as a proxy for the fluid-flow simulator for a fraction
of the computational cost. We demonstrate the efficacy of our proposed
method by means of a synthetic experiment. Finally, our framework is
extended to carbon sequestration forecasting, where we effectively use
the surrogate Fourier neural operator to forecast the
CO\textsubscript{2} plume in the future at near-zero additional cost.

\vspace*{-0.3cm}

\section{Introduction}\label{introduction}

\vspace*{-0.2cm}

Time-lapse seismic monitoring of CO\textsubscript{2} sequestration is
one of the most commonly used technologies to monitor
CO\textsubscript{2} dynamics in the Earth's subsurface through multiple
seismic surveys (vintages) \citep{lumley2001time}. Time-lapse seismic
has been used by carbon capture and storage (CCS) practitioners at
various storage sites
\citep{eiken2000seismic, ARTS2003347, chadwick2010quantitative, ringrose2013salah, FURRE20173916}.
The growth of the CO\textsubscript{2} plumes can be inferred by
time-lapse seismic imaging
\citep{ARTS2003347, ayeni2010target, kotsi2020time, yin2021SEGcts} or by
time-lapse full-waveform inversion
\citep{queisser2013full, yang2016time, kotsi2020uncertainty}.
Unfortunately, plain subtractions of time-lapse seismic images or
inversion results often contain unwanted artifacts due to noise or to
differences in acquisition \citep{oghenekohwo2017highly, zhou2021non},
which can potentially corrupt the often rather subtle time-lapse
differences due to changes in CO\textsubscript{2} concentration.

Over the years, several attempts have been made to mitigate this
challenge by improving the repeatability of time-lapse seismic,
including the forward and backward bootstrapping method
\citep{asnaashari2015time}, the double-difference method
\citep{watanabe2004differential, denli2009double, zhang2013double, yang2015double},
the central-difference method \citep{zhou2021central}, data assimilation
via Kalman filtering
\citep{li2014kalman, eikrem2019iterated, huang2020towards} and the joint
recovery model
\citep{oghenekohwo2016GEOPctl, wason2016GEOPctl, oghenekohwo2017EAGEitl, yin2021SEGcts}.
While these methods resulted in improvements in repeatability of
time-lapse seismic, they ignore the fact that the dynamics of
CO\textsubscript{2} plumes, to the leading order, adhere to two-phase
flow equations. Given physical properties of the two fluids (brine and
supercritical CO\textsubscript{2}) and the spatial porosity and
permeability distributions, these fluid-flow equations are capable of
predicting CO\textsubscript{2} concentration snapshots during and after
CO\textsubscript{2} injection. By coupling these fluid-flow equations,
via a rock physics model (the patchy saturation model
\citep{avseth2010quantitative}), to the wave equation,
\citet{li2020coupled} proposed an end-to-end inversion framework where
time-lapse seismic surveys are jointly inverted to yield estimates for
the spatial permeability distribution. Compared to the sequential
inversion \citep{hatab2021assessment, hatab2021assessing} and history
matching workflows \citep{oliver20214d}, estimates of the
CO\textsubscript{2} concentration in the coupled inversion are
regularized by fluid-flow physics. Because coupled inversion
\citep{li2020coupled} makes use of the fluid-flow equations, it offers a
framework capable of producing direct estimates for the permeability.
The latter can be used to generate improved predictions for the behavior
of CO\textsubscript{2} plumes. Despite the initially promising results
by \citet{li2020coupled} on synthetic experiments, the downside of their
proposed coupled framework is the increased complexity that comes with
including partial differential equations (PDEs) for the fluid-flow as
constraints. Aside from the need to compute sensitivities of solutions
of the fluid-flow equations, solving these PDE can be computationally
expensive \citep{settgast2018geosx, rasmussen2021open}.

To address this challenge, we propose to replace solvers for the
fluid-flow PDEs by Fourier neural operators
\citep[FNOs,][]{li2020fourier}. After training on a representative
dataset, FNOs are capable of producing CO\textsubscript{2} plume
snapshots quickly \citep{wen2021u, zhang2022fourier} while automatic
differentiation (AD) gives us easy and fast access to the gradient with
respect to the input. As such, trained FNOs can be considered as a
data-driven surrogate for the computationally expensive fluid-flow
simulations, making them a suitable candidate for the proposed coupled
inversion framework that calls for multiple fluid-flow simulations
including calculation of the gradient.

This extended abstract is organized as follows. First, we discuss the
coupled inversion framework for seismic monitoring of geological carbon
storage. Second, we introduce FNOs as surrogates for fluid-flow
simulations. Next, we present the learned FNO-based coupled inversion
framework where the fluid-flow solver is replaced by a pre-trained FNO.
Finally, we verify the efficacy of the learned coupled inversion
framework through a synthetic experiment. We further show that the
trained FNO can forecast the growth of the CO\textsubscript{2} plume in
the future with the inverted permeability model.

\vspace*{-0.3cm}

\section{Coupled inversion framework}\label{coupled-inversion-framework}

\vspace*{-0.2cm}

Our goal is to estimate the past, current, and future behavior of
CO\textsubscript{2} plumes from available time-lapse seismic data. To
achieve this goal, we consider the coupled inversion framework proposed
by \citet{li2020coupled}. In this framework, three types of physics are
integrated, namely fluid-flow, rock, and wave physics. The
CO\textsubscript{2} plume dynamics are modeled by two-phase flow
equations \citep{pruess2011numerical}, which we represent as the
following mapping:
\begin{equation}
\mathbf{K}\mapsto\mathbf{c}=\mathcal{S}(\mathbf{K})\quad\text{where} \quad \mathbf{c}=[\mathbf{c}_1,\mathbf{c}_2,\ldots,\mathbf{c}_{n_v}],
\label{fluidflow}
\end{equation}
 where the vectors $\mathbf{c}_i,\,i=1\ldots n_v$ are the discretized
CO\textsubscript{2} concentration snapshots for each vintage. In this
mapping, $\mathcal{S}$ represents two-phase flow simulations that given
the permeability, $\mathbf{K}$, models the CO\textsubscript{2}
concentration as a spatial function over $n_v$ consecutive times. Then,
the patchy saturation model \citep{avseth2010quantitative} maps the
$n_v$ snapshots of the CO\textsubscript{2} concentration to seismic
wavespeeds---i.e., we introduce for each vintage ($i=1\ldots n_v$) the
following mapping:
\begin{equation}
\mathbf{c}_i\mapsto\mathbf{v}_i = \mathcal{R}(\mathbf{c}_i) \quad \text{for} \quad i=1, 2, \ldots, n_v,
\label{patchy}
\end{equation}
 where $\mathcal{R}$ represents the rock physics model and
$\mathbf{v}_i,\, i=1\cdots n_v$ the snapshots of the acoustic wavespeed.
To connect these wavespeed snapshots to the seismic data vintages, we
introduce the mapping:
\begin{equation}
\mathbf{v}_i\mapsto\mathbf{d}_i = \mathcal{F}_i(\mathbf{v}_i) \quad \text{for} \quad i=1, 2, \ldots, n_v,
\label{seismic}
\end{equation}
 where $\mathcal{F}_i$ is the wave modeling operator for vintage $i$
that generates the corresponding seismic dataset $\mathbf{d}_i$ given
the velocity model $\mathbf{v}_i$
\citep{tarantola1984inversion, virieux2009overview}. To arrive at the
end-to-end formulation linking multi-vintage seismic data to the
permeability, we finally compose (denoted by the $\circ$ symbol) these
three mappings yielding the following minimization problem:
\begin{equation}
\begin{split}
\underset{\mathbf{K}}{\operatorname{minimize}} \quad \frac{1}{2}\|\mathcal{F}\circ\mathcal{R}\circ\mathcal{S}(\mathbf{K})-\mathbf{d}\|_2^2 \\
\text{where} \quad \mathbf{d}=[\mathbf{d}_1,\mathbf{d}_2,\ldots,\mathbf{d}_{n_v}],
\end{split}
\label{couple}
\end{equation}
 where $\mathbf{d}$ represents the $n_v$ vintages of the time-lapse
data. While the optimization problem~\ref{couple} offers a unique
formulation where time-lapse seismic data are linked to the
permeability, its minimization is complex since it entails nested
application of the adjoint-state method \citep{plessix2006review}
involving computationally expensive forward simulations of both wave and
fluid-flow physics. To simplify the formulation and to drastically
reduce computational cost of minimization problem~\ref{couple}, we
propose to replace the fluid-flow solves by a trained FNO.

\vspace*{-0.3cm}

\section{Fourier neural operators}\label{fourier-neural-operators}

\vspace*{-0.2cm}

There exists a growing literature on solving numerical PDEs via learned
data-driven approaches involving neural networks
\citep{lu2019deeponet, raissi2019physics, kochkov2021machine, karniadakis2021physics}.
After incurring initial training costs, neural networks have been shown
to to provide faster alternatives to numerical PDE simulations.
Recently, FNOs \citep{li2020fourier, li2021physics} have emerged as a
powerful technique to approximate the solution operator of parametric
PDEs. After training, FNOs can generate approximation solutions of PDEs
from the coefficients orders of magnitude faster than numerical solvers
\citep{li2020fourier}. This means that computational costs, which
consist of generating training pairs coefficient (permeability
$\mathbf{K}$) and solution (CO\textsubscript{2} concentration
$\mathbf{c}$) and training the network, are sustained upfront. This
front loading of computations leads to a drastic reduction in simulation
time during the inversion---i.e., minimization of problem~\ref{couple}.
We refer to the existing literature \citep{wen2021u, zhang2022fourier}
for details on how to train FNOs to approximately map permeability
models to the time evolution of CO\textsubscript{2} plumes. In this
abstract, we assume to have
$\mathcal{S}_{\boldsymbol{\theta}}(\mathbf{K})\approx \mathcal{S}(\mathbf{K})$
for the permeability drawn from a certain distribution. Here,
$\mathcal{S}_{\boldsymbol{\theta}}(\cdot)$ denotes the approximate map,
which depends on the learned FNO weights ${\boldsymbol{\theta}}$. Given
an unseen spatial distribution for the permeability $\mathbf{K}$, the
trained FNO can instantaneously produce the time-dependant
CO\textsubscript{2} concentration by forward evaluation of the FNO as
$\mathcal{S}_{\boldsymbol{\theta}}(\mathbf{K})$
\citep{li2020fourier, wen2021u, zhang2022fourier}. In addition, AD gives
us access to the gradient with respect to FNO's input (the permeability
$\mathbf{K}$) that can be used for inversion. By virtue of these
capabilities, the proposed approximation by FNOs can be used as a
surrogate for the fluid-flow solver in the end-to-end formulation of
problem~\ref{couple}.

\vspace*{-0.3cm}

\section{Forecast via learned coupled
inversion}\label{forecast-via-learned-coupled-inversion}

\vspace*{-0.2cm}

\begin{figure}
\centering
\includegraphics[width=1.000\hsize]{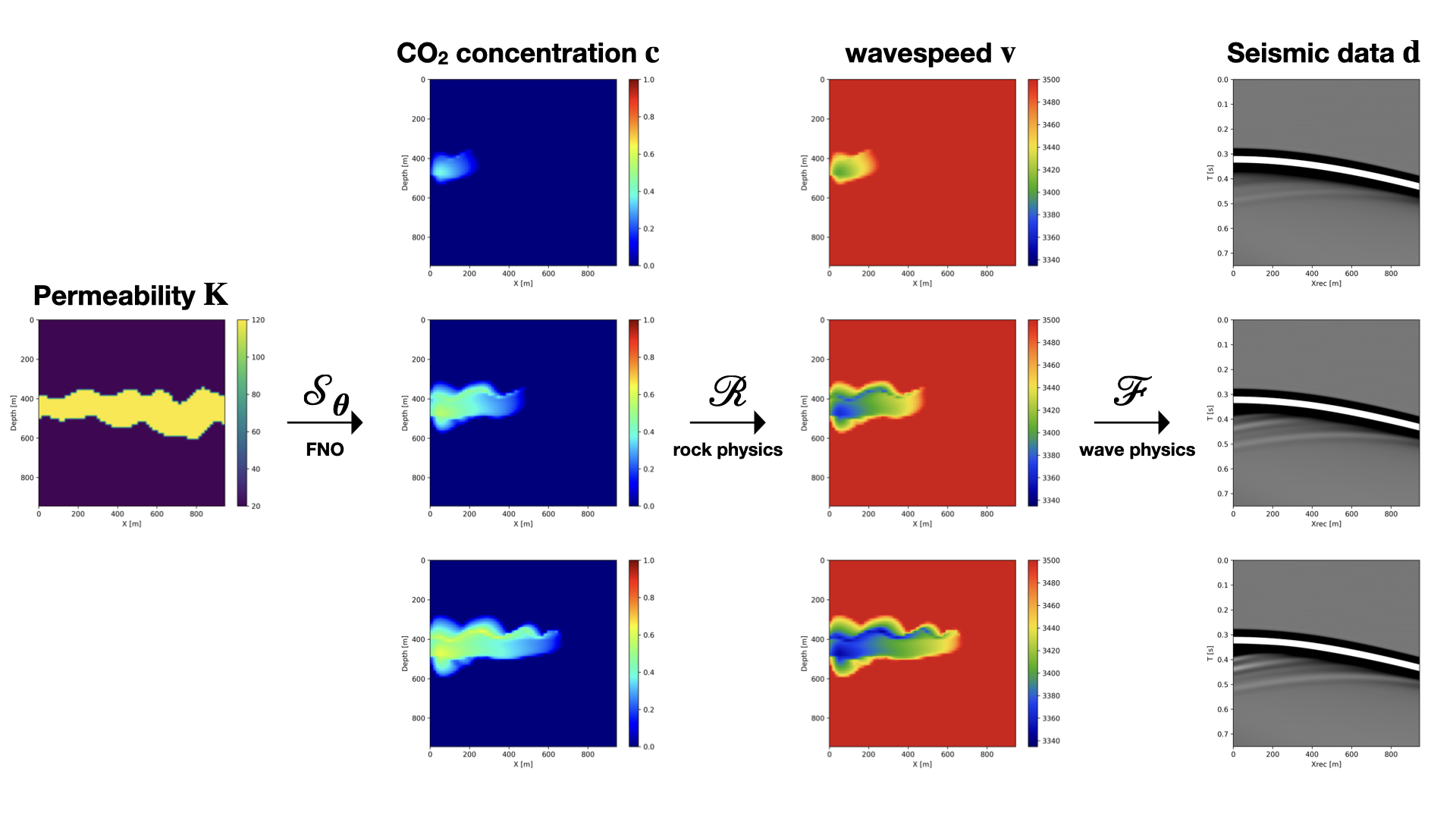}
\caption{Learned coupled inversion framework, which contains three
modules: pre-trained FNO, rock physics model, and wave physics model.
Given time-lapse seismic datasets, we estimate the intrinsic
permeability via end-to-end inversion.}\label{fig:framework}
\end{figure}

While the end-to-end formulation (shown in Figure~\ref{fig:framework})
provides access to estimates of the permeability, CO\textsubscript{2}
plume forecasting \citep{wen2021towards} is our main objective because
they offer guarantees that the CO\textsubscript{2} plume is progressing
as planned. To meet our goal of CO\textsubscript{2} plume forecasting,
the proposed learned and coupled formulation offers several distinct
advantages. First, the coupled formulation uses information from all
collected time-lapse vintages to arrive at estimates for the
permeability itself and past and current behavior of the
CO\textsubscript{2} plume. The fact that the two-phase flow equations
act as a regularizer leads to improved estimates for the
CO\textsubscript{2} plume \citep{li2020coupled}. Second, the use of FNOs
reduces the computational cost \citep{li2020fourier, wen2021u}, which
potentially enables uncertainty quantification and risk management of
the growth of the CO\textsubscript{2} plume in the future. Third, the
coupled framework provides access to estimates for the permeability that
can be used to forecast future behavior of CO\textsubscript{2} plumes.
Given the estimated permeability model from the learned coupled
inversion and an FNO trained on the time range of CCS projects, the
current and future CO\textsubscript{2} concentration snapshots can all
be generated by forward evaluation of the FNO. The forecast of
CO\textsubscript{2} plume in the future can help the practitioners to
detect potential failing scenarios in the early period of the CCS
project, such as CO\textsubscript{2} leaking through fractures in the
seal \citep{ringrose2020store}.

\vspace*{-0.3cm}

\section{Numerical experiments}\label{numerical-experiments}

\vspace*{-0.2cm}

By means of a synthetic case study, we validate the performance of our
FNO-based learned coupled inversion framework to invert for the
permeability from time-lapse seismic data. We then show that we can use
the estimated permeability to forecast the evolution of the
CO\textsubscript{2} plume in the future. We begin by describing how we
train the FNO to learn the two-phase flow physics.

\vspace*{-0.1cm}

\subsection{Training setup}\label{training-setup}

\vspace*{-0.1cm}

We create $1000$ pairs of permeability and time evolution of
CO\textsubscript{2} concentration to form the training set. The size of
the permeability is $64 \times 64$ with a grid spacing of
$15\,\mathrm{m}$ in both vertical and horizontal directions. Each
permeability model is $20\,$ millidarcies (md) everywhere except for a
high permeability channel with $120\,\mathrm{md}$ in the central area of
the model, of which the random curving boundaries are generated by a
Gaussian process \citep{bishop2006pattern}. An example of the
permeability model in shown in Figure~\ref{fig:k-true}. For fluid-flow
simulation, we add an injection well that injects supercritical
CO\textsubscript{2} (with density $501.9\,\mathrm{kg}/\mathrm{m}^3$) on
the left-hand side of the model, and a production well that produces
brine (with density $1053.0\,\mathrm{kg}/\mathrm{m}^3$) on the
right-hand side of the model. We assume the porosity of the reservoir is
fixed to $25\%$ homogeneously. The time evolution of the
CO\textsubscript{2} concentration is modeled with
\href{https://github.com/lidongzh/FwiFlow.jl}{FwiFlow.jl}
\citep{li2020coupled} over a period of $1000$ days with a time step of
$20$ days. This numerical simulation creates $51$ snapshots in total for
each permeability model. We form the training dataset, where each sample
is a pair of the permeability model and corresponding $51$ snapshots of
CO\textsubscript{2} concentration. We train an FNO that maps the input
permeability $\mathbf{K}(x,z,t)$ to the output CO\textsubscript{2}
concentration $\mathbf{c}(x,z,t)$. We follow the original implementation
of FNOs \url{https://github.com/zongyi-li/fourier_neural_operator}
\citep{li2020fourier} and re-implement the FNO in Julia in order to
integrate with other software modules in the learned coupled inversion
framework.

\vspace*{-0.1cm}

\subsection{Learned coupled inversion from time-lapse seismic
data}\label{learned-coupled-inversion-from-time-lapse-seismic-data}

\vspace*{-0.1cm}

\begin{figure}
\centering
\subfloat[\label{fig:k-true}]{\includegraphics[width=0.333\hsize]{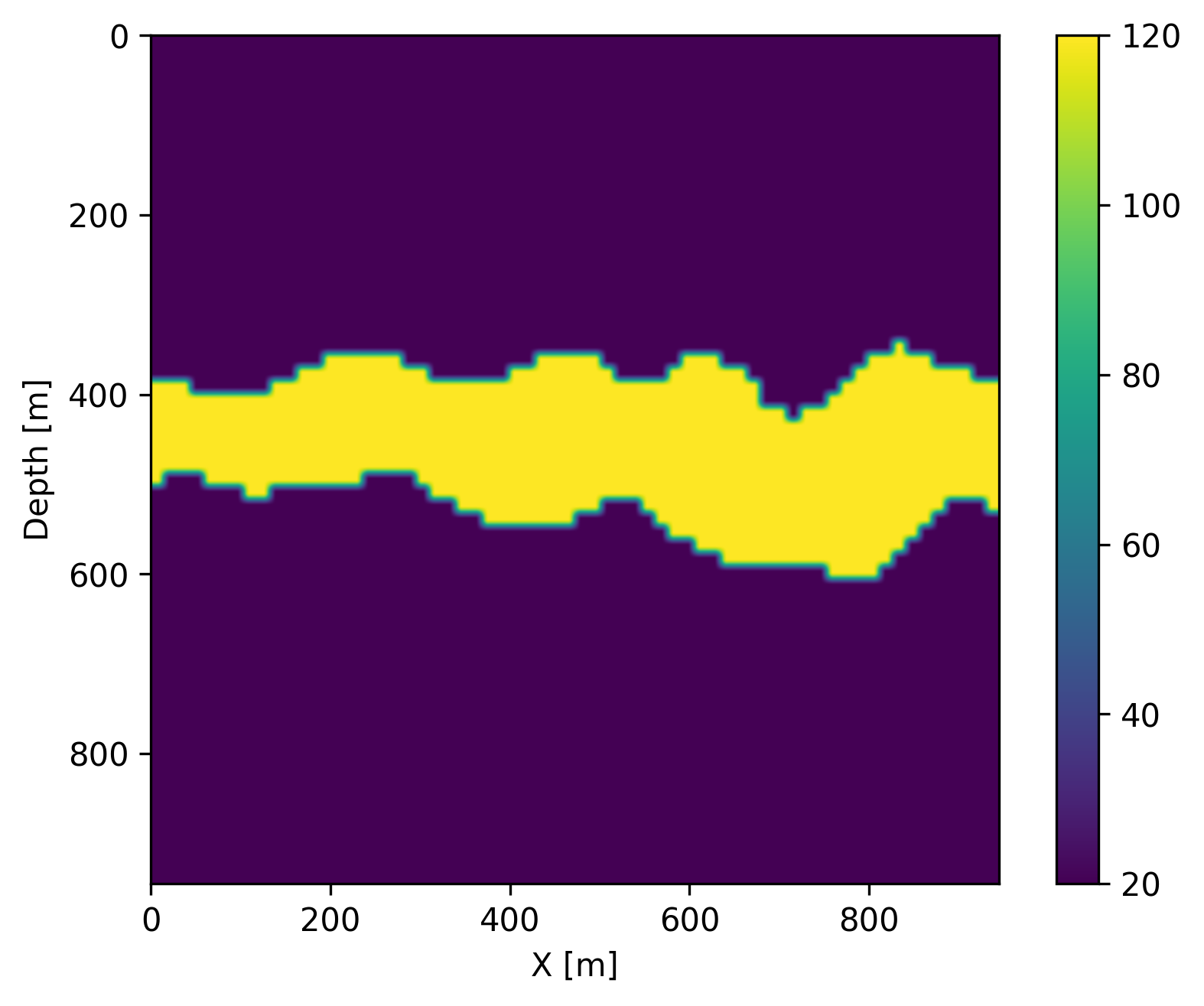}}
\subfloat[\label{fig:k-init}]{\includegraphics[width=0.333\hsize]{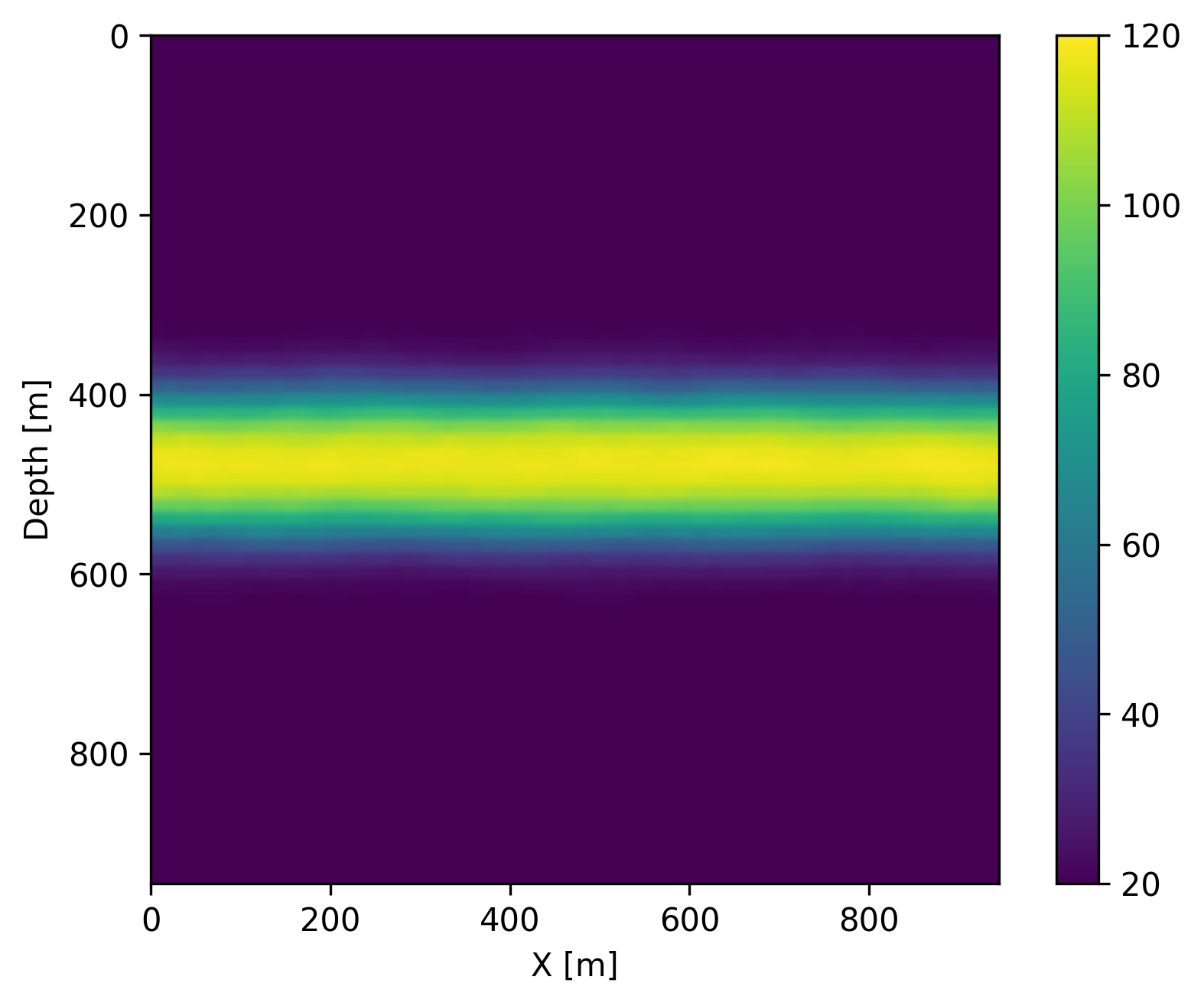}}
\subfloat[\label{fig:k-inv}]{\includegraphics[width=0.333\hsize]{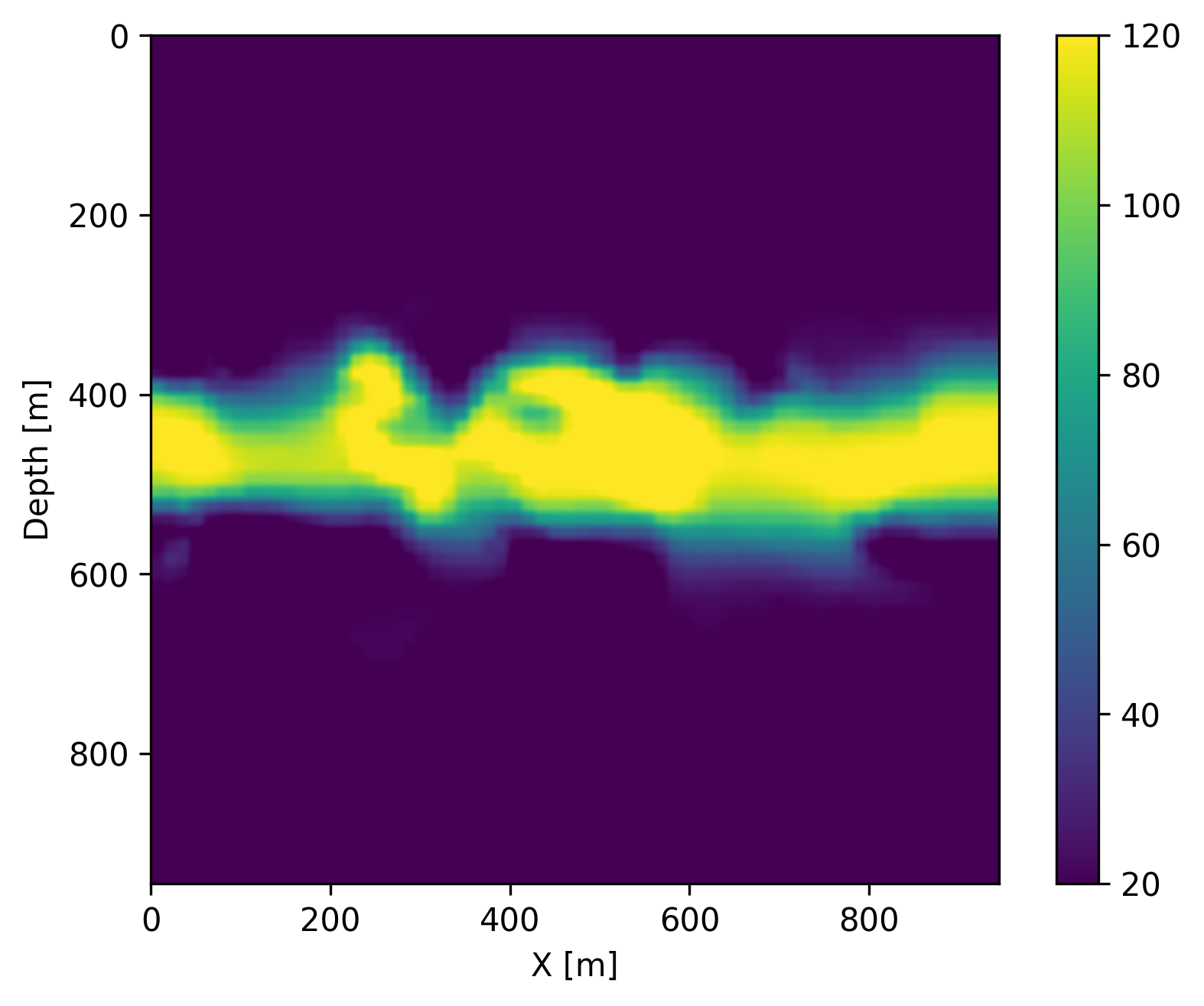}}
\caption{Learned coupled inversion from $11$ seismic surveys with FNO as
a surrogate. (a) Unseen ground truth permeability test sample. (b)
Initial permeability. (c) Inverted permeability via $100$ projected
gradient descent iterations.}\label{fig:inv}
\end{figure}

\begin{figure}
\centering
\subfloat[\label{fig:recover-1}]{\includegraphics[width=0.330\hsize]{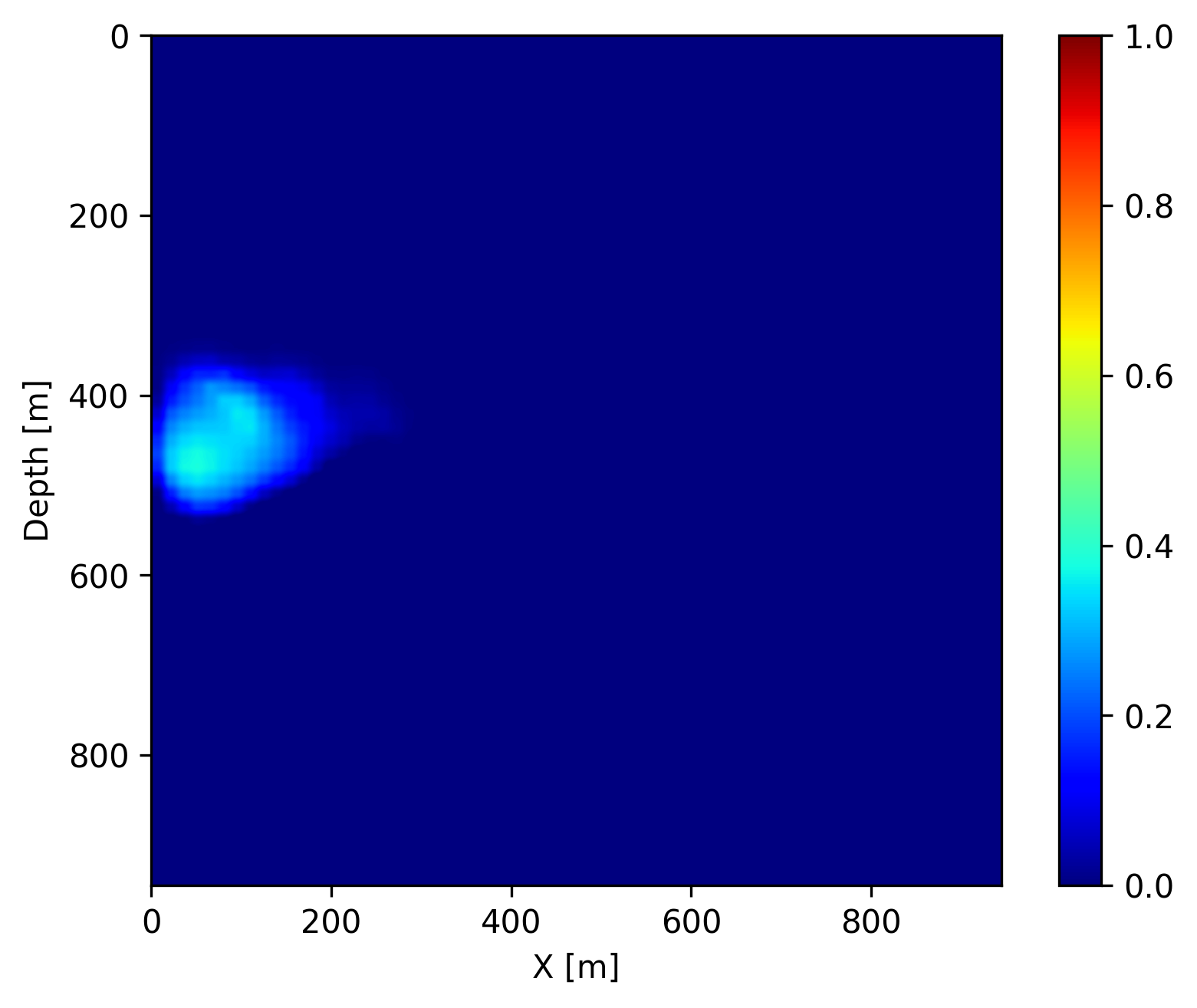}}
\subfloat[\label{fig:recover-2}]{\includegraphics[width=0.330\hsize]{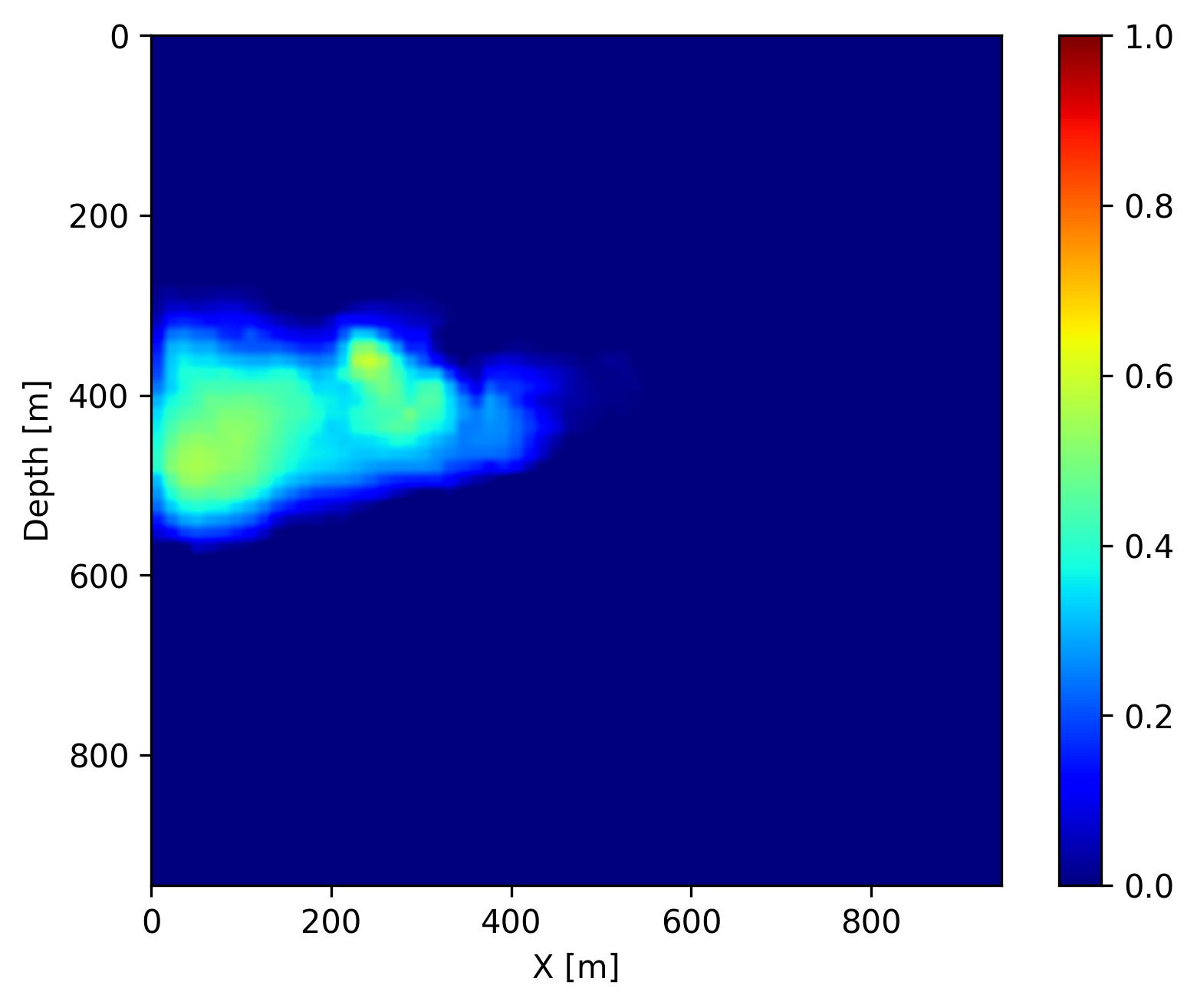}}
\subfloat[\label{fig:recover-3}]{\includegraphics[width=0.330\hsize]{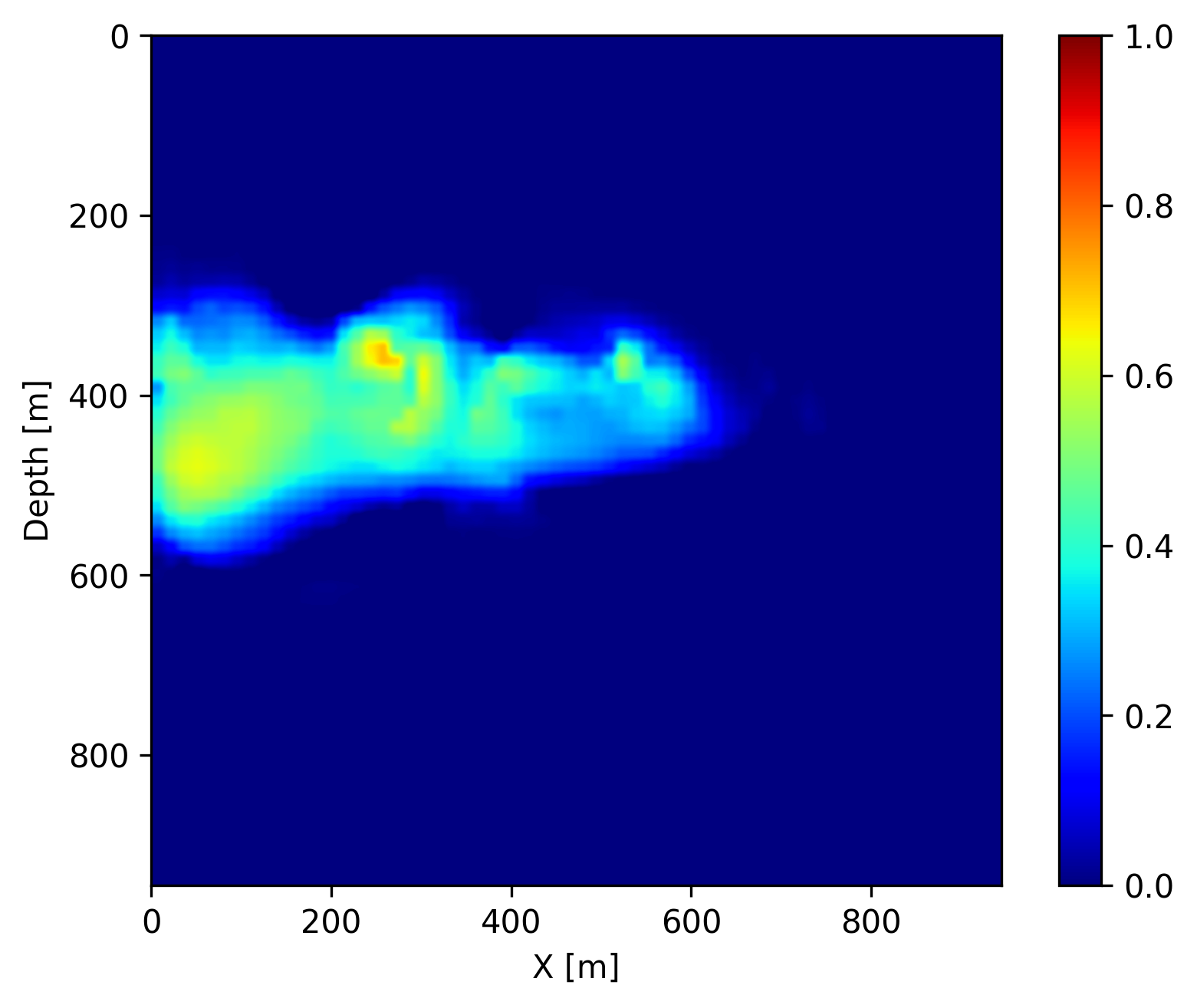}}
\\
\subfloat[\label{fig:true-1}]{\includegraphics[width=0.330\hsize]{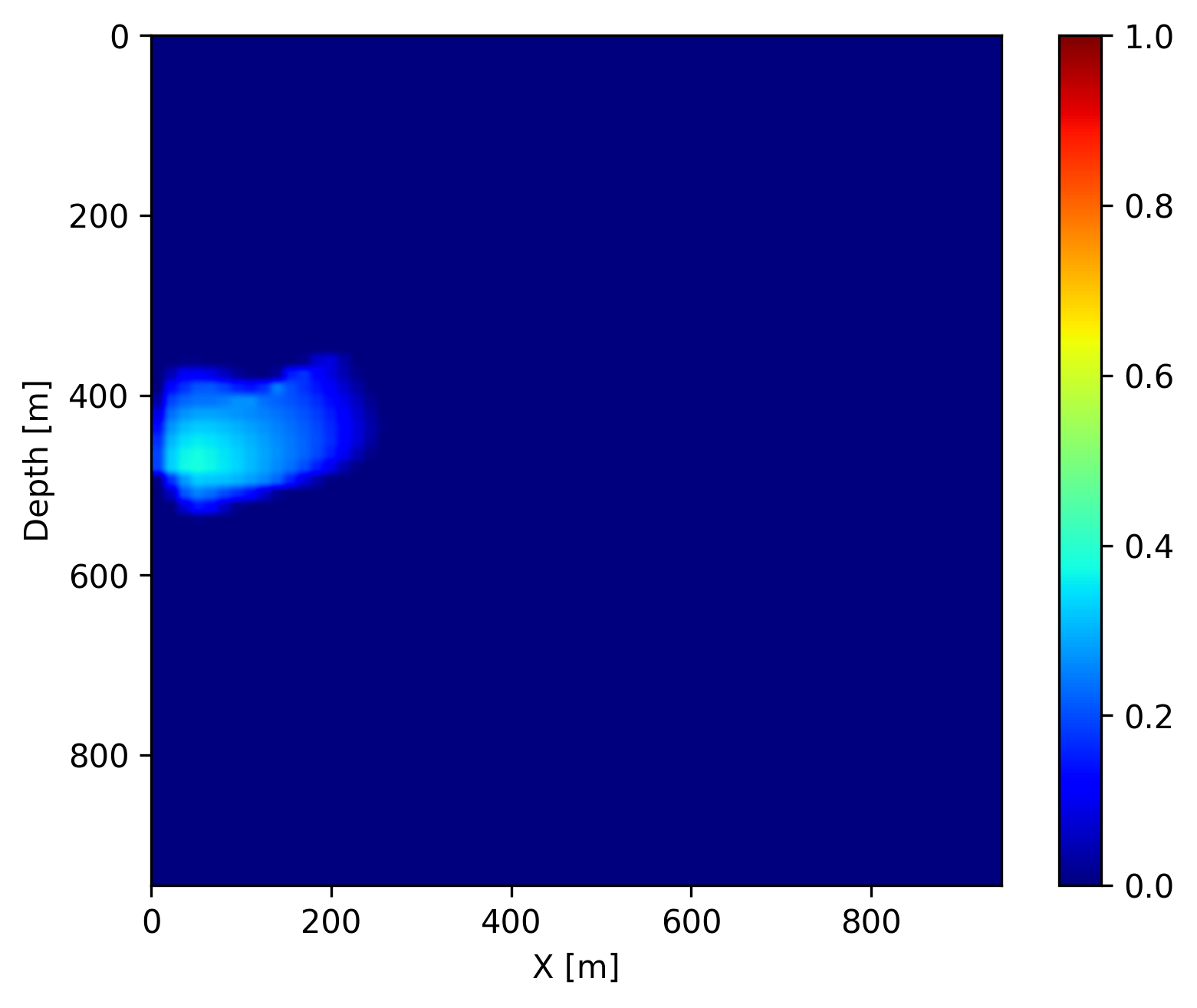}}
\subfloat[\label{fig:true-2}]{\includegraphics[width=0.330\hsize]{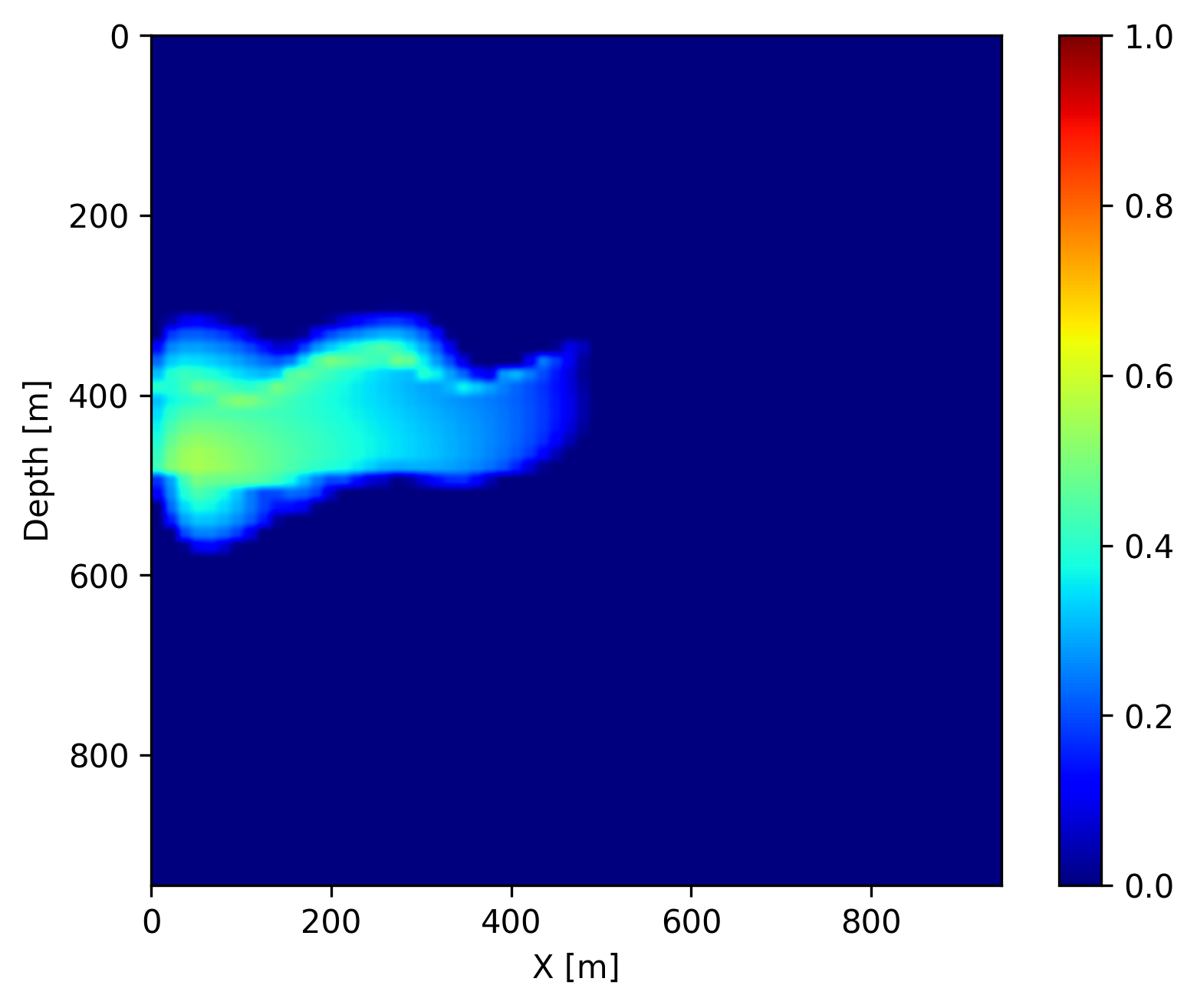}}
\subfloat[\label{fig:true-3}]{\includegraphics[width=0.330\hsize]{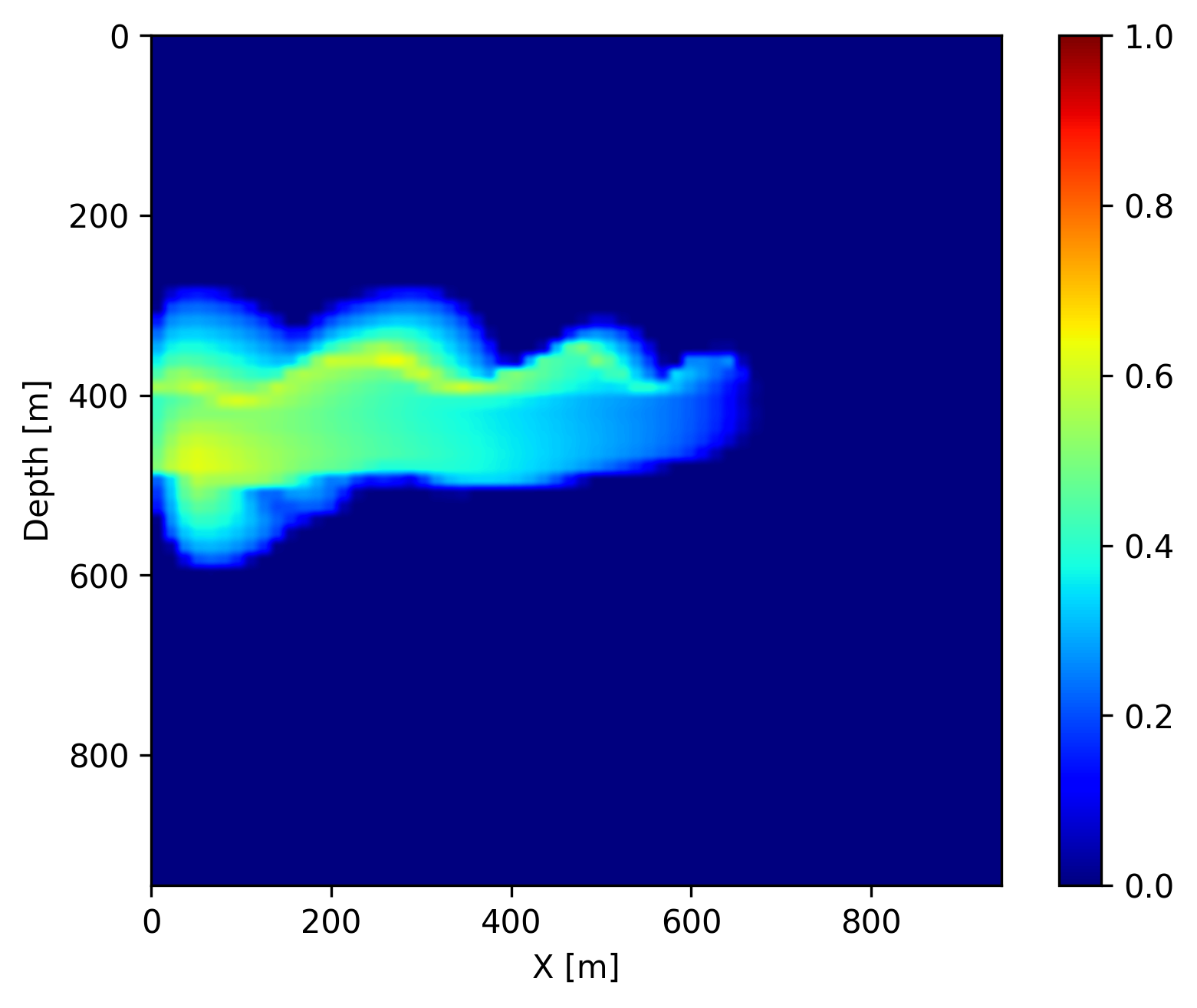}}
\\
\subfloat[\label{fig:diff-1}]{\includegraphics[width=0.330\hsize]{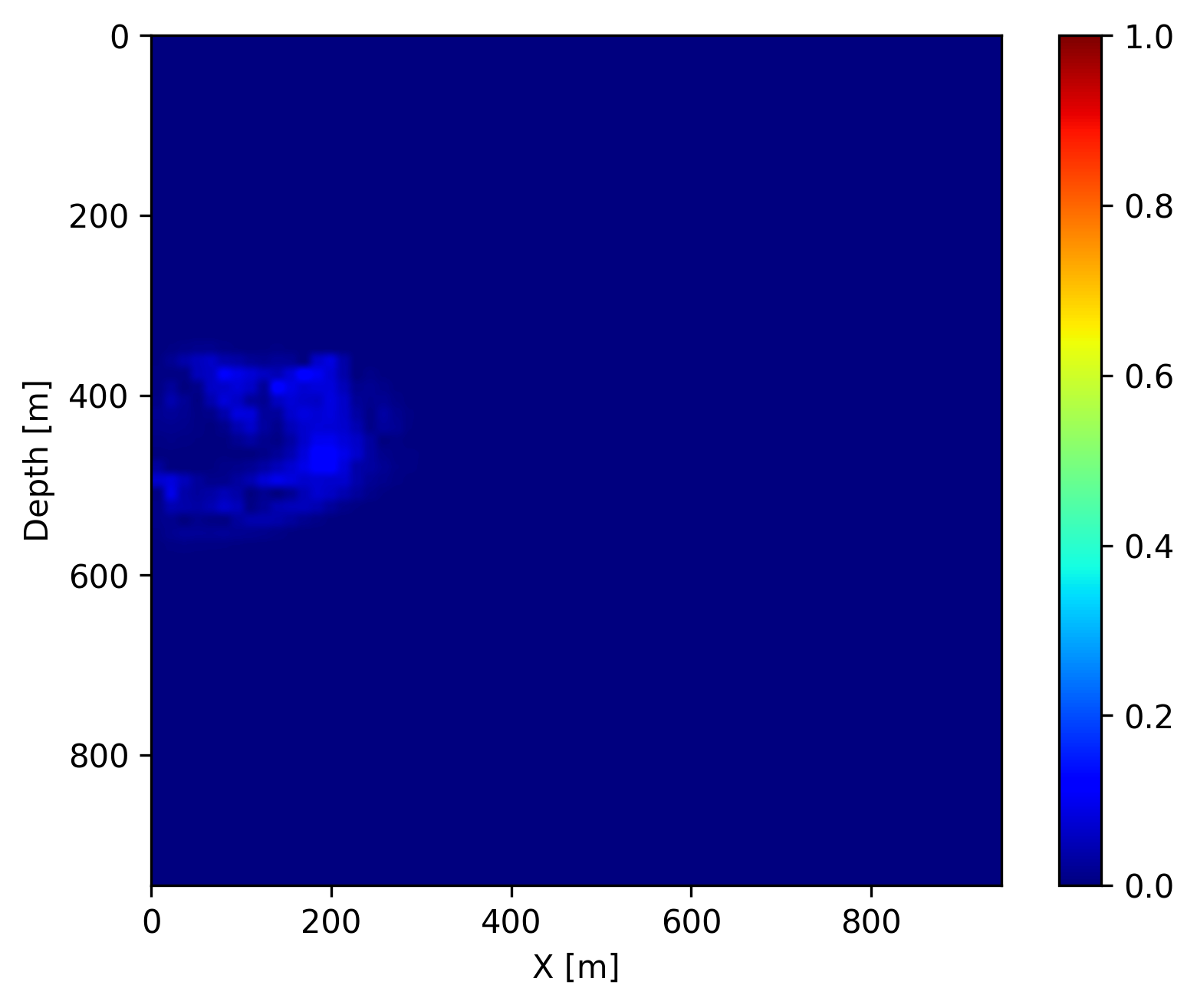}}
\subfloat[\label{fig:diff-2}]{\includegraphics[width=0.330\hsize]{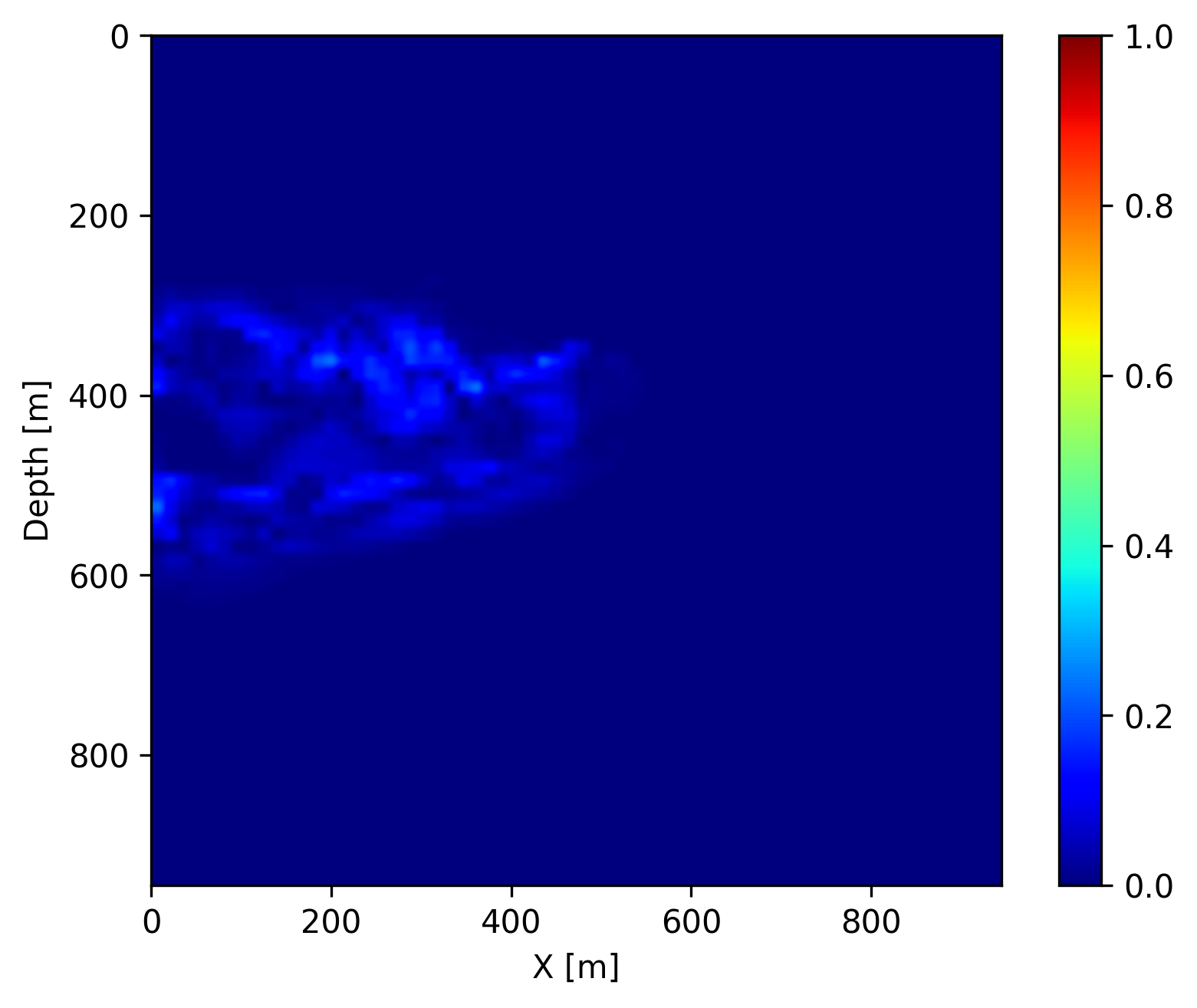}}
\subfloat[\label{fig:diff-3}]{\includegraphics[width=0.330\hsize]{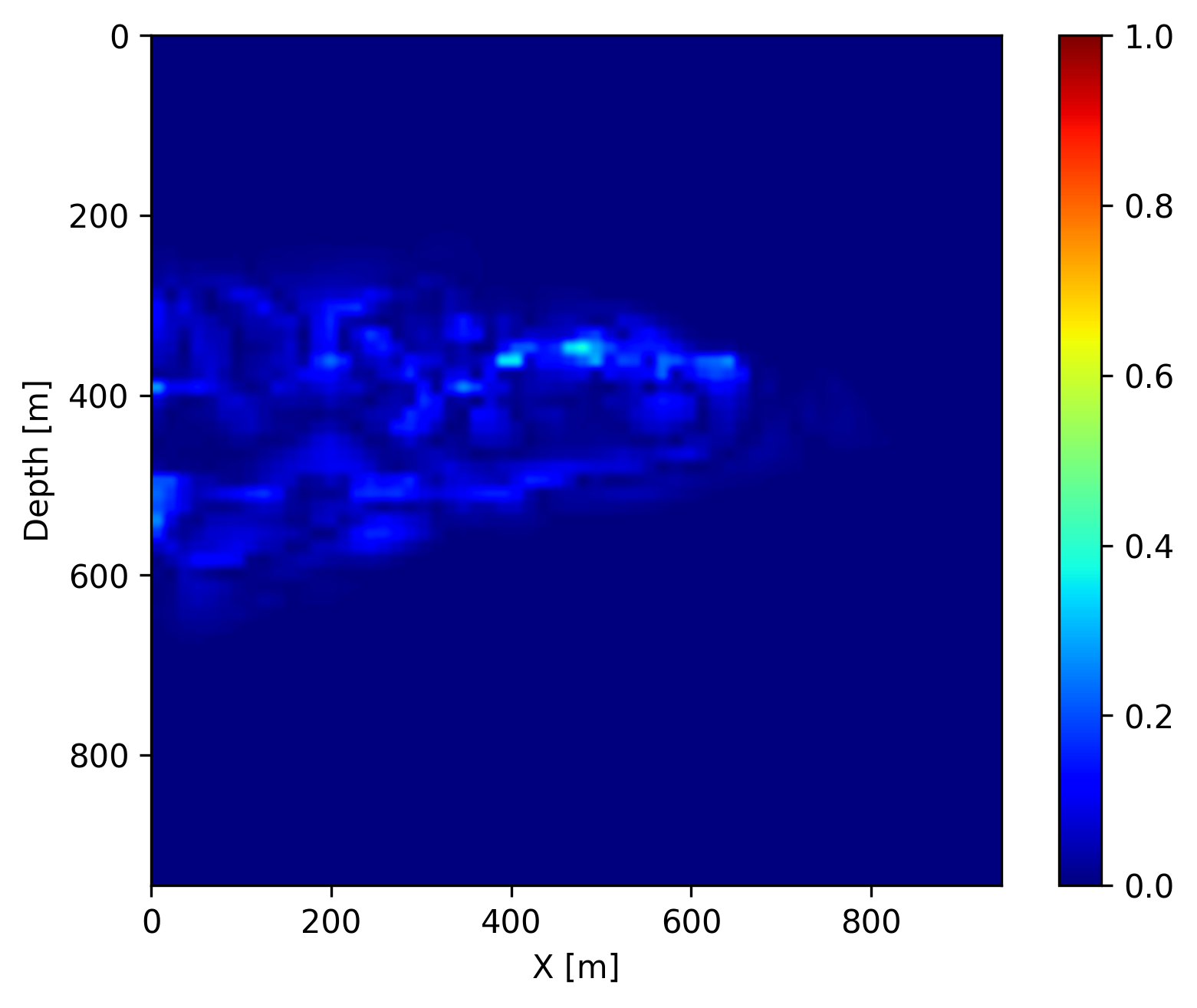}}
\caption{(a)(b)(c) Recovered snapshots of CO\textsubscript{2}
concentration at $40$, $160$, and $280$ days after injection. (d)(e)(f)
Ground truth snapshots of CO\textsubscript{2} concentration. (g)(h)(i)
Difference plotted in the same scale.}\label{fig:recover}
\end{figure}

After training, we show the performance of the learned coupled inversion
framework described in Figure~\ref{fig:framework}. During testing, we
draw an unseen permeability sample, shown in Figure~\ref{fig:k-true},
and generate $11$ early snapshots of CO\textsubscript{2} concentration
at every $40$th day using the numerical solvers. The snapshots at day
$40$, $160$ and $280$ are shown in Figure~\ref{fig:true-1},
\ref{fig:true-2}, \ref{fig:true-3}, respectively. Since these are the
early snapshots, the CO\textsubscript{2} plume does not reach the entire
high permeability channel from the left to the right. We then convert
these CO\textsubscript{2} concentration snapshots to the time-varying
velocity models via the patchy saturation model
\citep{avseth2010quantitative}, and generate $11$ cross-well seismic
surveys on these $11$ velocity models. The wave physics is modeled with
\href{https://github.com/slimgroup/JUDI.jl}{JUDI.jl}
\citep{witte2018alf, mathias_louboutin_2022_5893940}, which uses the
highly-optimized matrix-free wave propagators of
\href{https://www.devitoproject.org}{Devito}
\citep{louboutin2018dae, luporini2020architecture, fabio_luporini_2022_6108644}.
Each seismic survey contains $32$ active sources in a borehole on the
left-hand side of the model, and $960$ receivers in a borehole on the
right-hand side of the model. We then invert for the permeability from
the time-lapse seismic data via the learned coupled inversion framework
in Figure~\ref{fig:framework}. We start our initial guess as an average
of the samples in the training set (a blurred channel shown in
Figure~\ref{fig:k-init}), and iteratively solve for the permeability by
projected gradient descent with a box constraint on the permeability
between $10\,\mathrm{md}$ and $130\,\mathrm{md}$. At each iteration, we
compute the seismic data misfit for only four shot records in each
time-lapse survey. This reduces the cost of wave physics as only one
eighth of the shot records are used during the forward and gradient
evaluation \citep{Li11TRfrfwi, van2013fast}. We use the backtracking
linear search algorithm \citep{stanimirovic2010accelerated} to choose
the step length accordingly. We use
\href{https://github.com/slimgroup/SetIntersectionProjection.jl}{SetIntersectionProjection.jl}
\citep{peters2019algorithms, bas_peters_2021_5203700} for box constraint
projection. After $120$ iterations ($15$ data passes on the entire
time-lapse dataset), the inverted permeability is shown in
Figure~\ref{fig:k-inv}. Since the CO\textsubscript{2} plume grows mostly
at the left part of the channel (near the injection well) in these early
snapshots, some part of the permeability model is in the null space thus
difficult to recover exactly. However, we can see that the learned
coupled inversion is able to approximately estimate the high
permeability channel and to delineate curvatures especially at the upper
boundary. Next, we demonstrate that this estimate is already remarkably
accurate to recover the shape of the CO\textsubscript{2} plume and
forecast the growth of the CO\textsubscript{2} plume in the future.

In addition to recovering the permeability, we are interested in
CO\textsubscript{2} concentration recovery as it indicates the growing
progress of the CO\textsubscript{2} plume. In Figure~\ref{fig:recover},
we show the recovered CO\textsubscript{2} concentration snapshots at day
$40$, $160$ and $280$, which are acquired by a forward evaluation of FNO
on the inverted permeability. We juxtapose the recovered concentrations
snapshots with the ground truth and the differences for easy
visualization. We observe that our predicted CO\textsubscript{2}
concentration snapshots fit the ground truth accurately with very few
artifacts near the boundaries of the CO\textsubscript{2} plume. This
demonstrates that the proposed learned coupled inversion framework can
be used successfully for carbon storage monitoring.

\begin{figure}
\centering
\subfloat[\label{fig:forecast-4}]{\includegraphics[width=0.330\hsize]{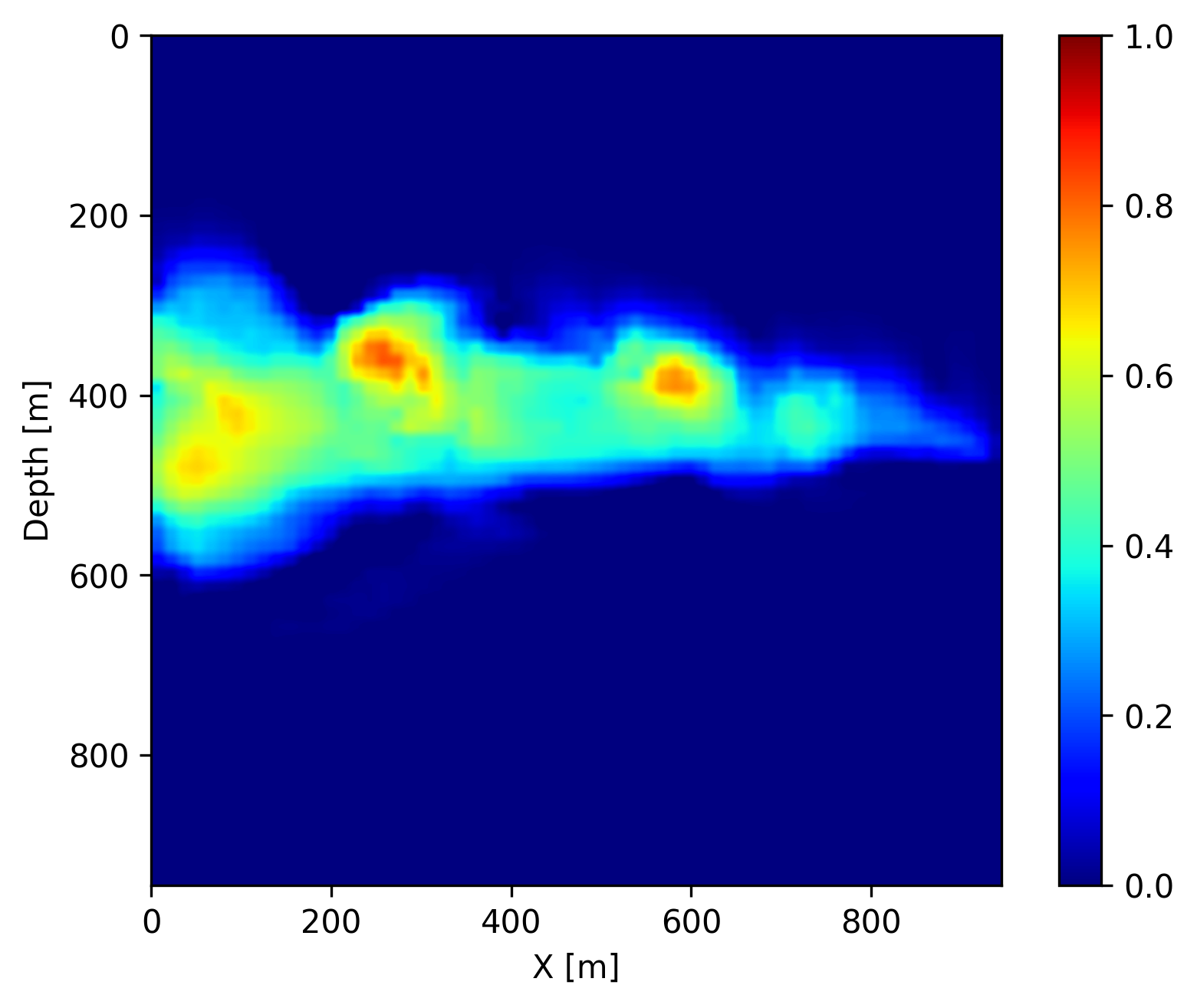}}
\subfloat[\label{fig:forecast-5}]{\includegraphics[width=0.330\hsize]{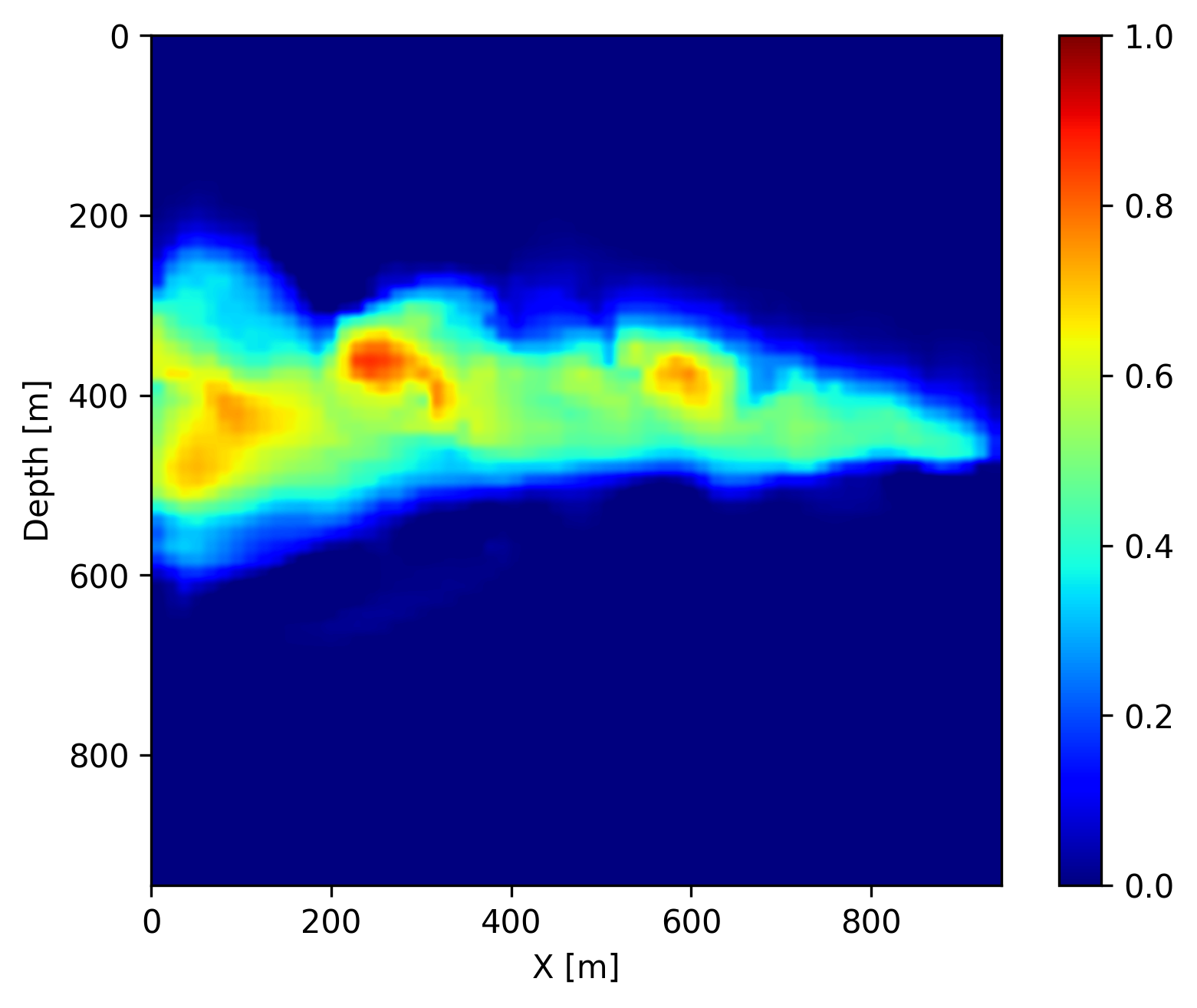}}
\subfloat[\label{fig:forecast-6}]{\includegraphics[width=0.330\hsize]{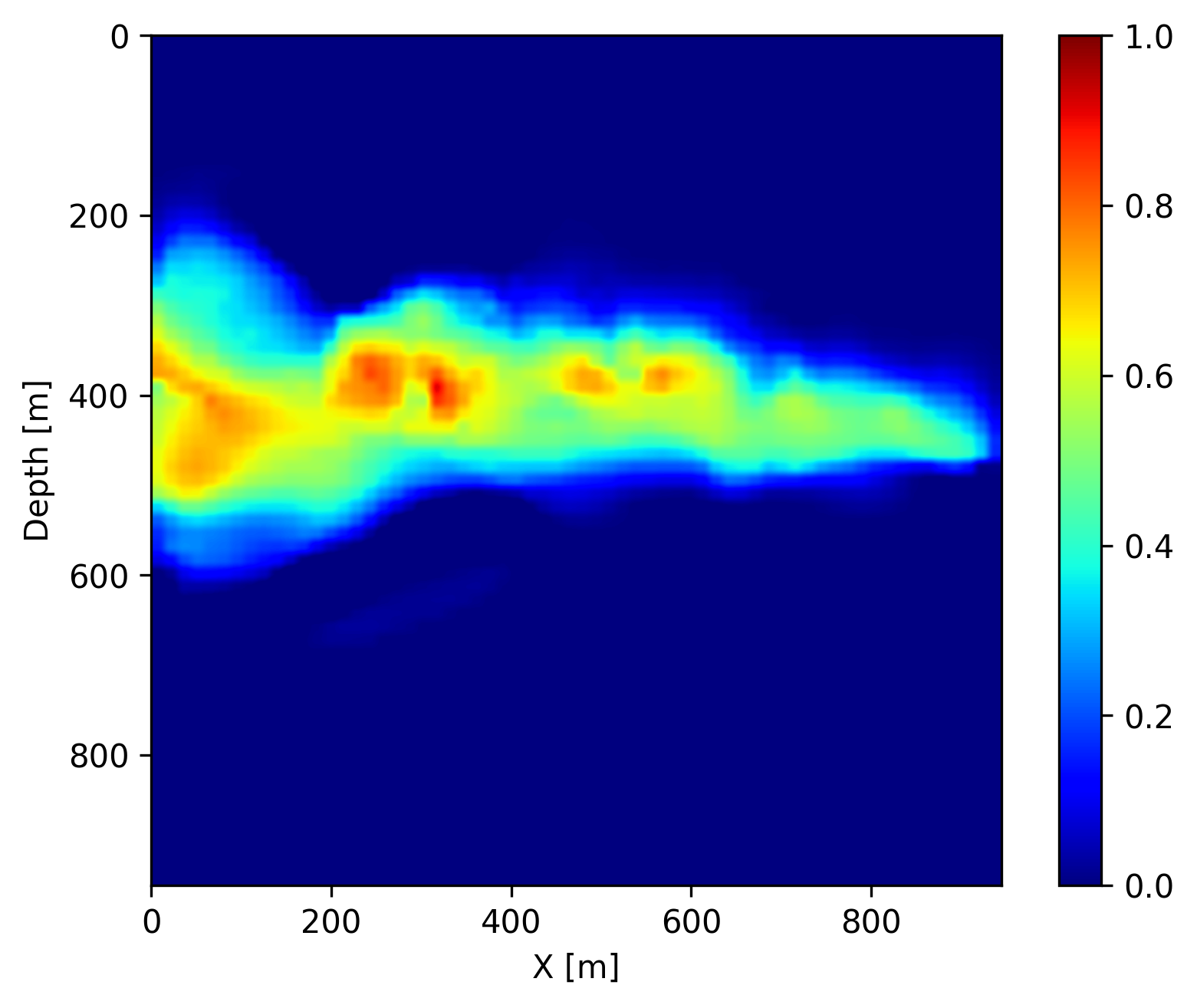}}
\\
\subfloat[\label{fig:true-4}]{\includegraphics[width=0.330\hsize]{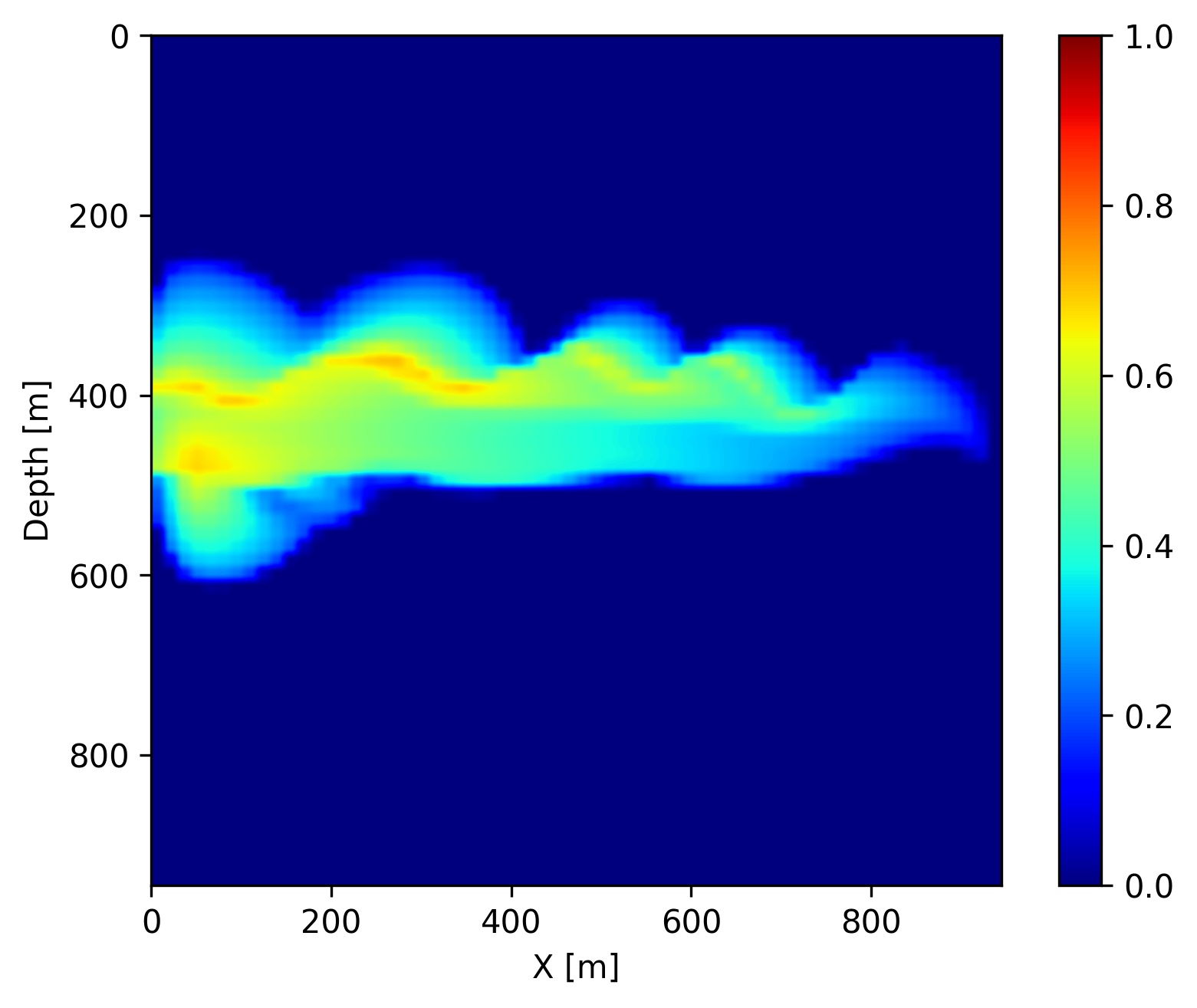}}
\subfloat[\label{fig:true-5}]{\includegraphics[width=0.330\hsize]{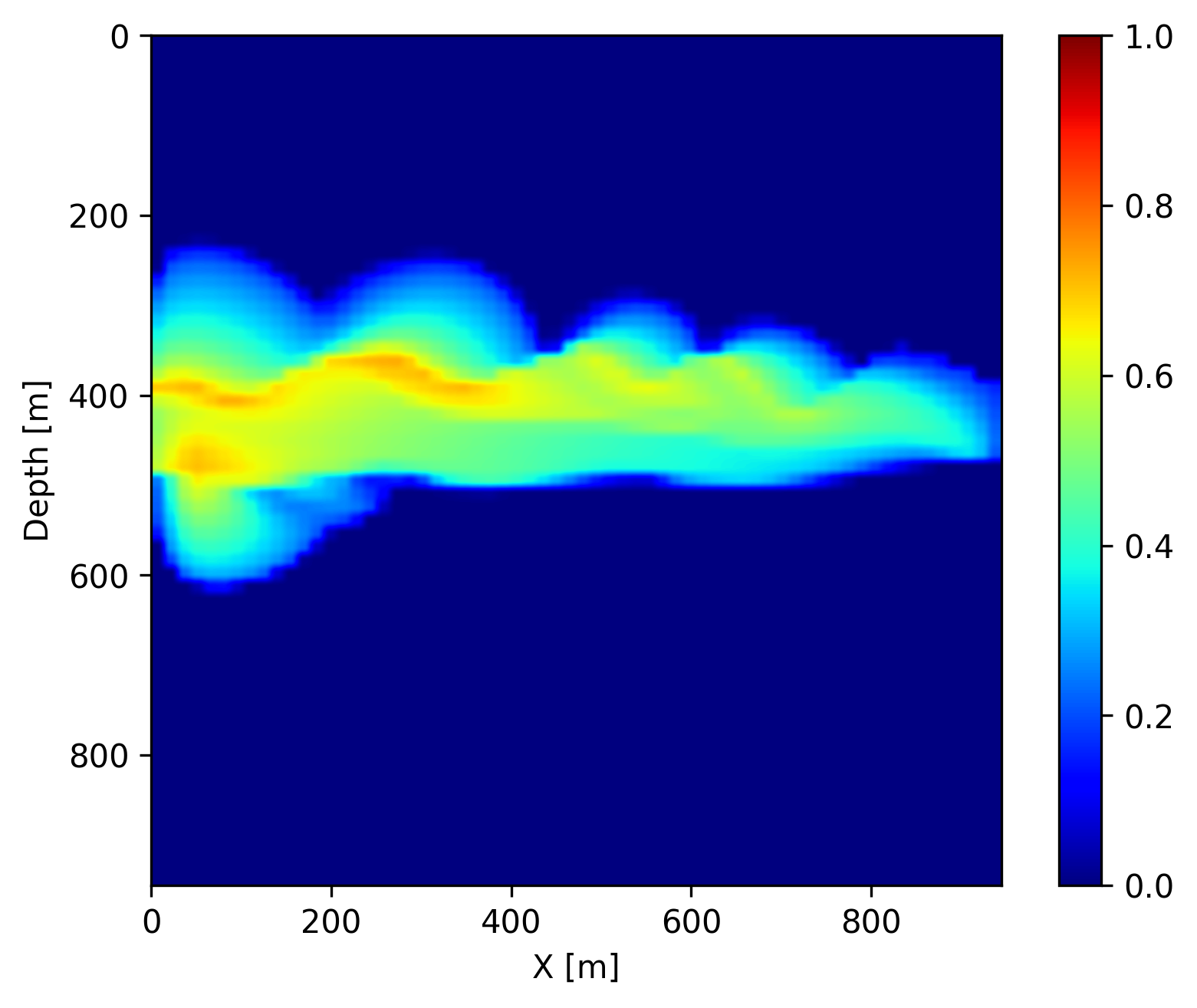}}
\subfloat[\label{fig:true-6}]{\includegraphics[width=0.330\hsize]{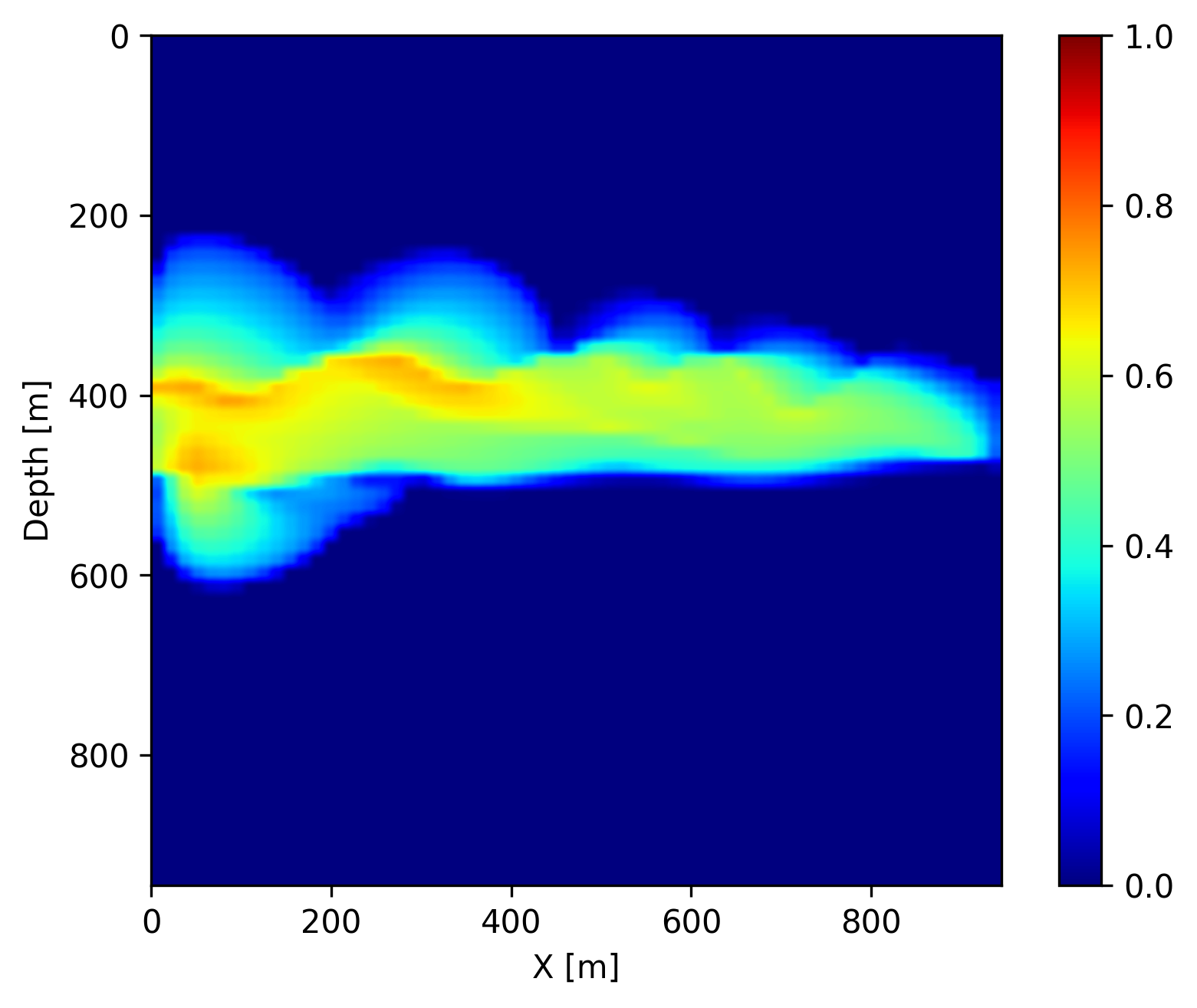}}
\\
\subfloat[\label{fig:diff-4}]{\includegraphics[width=0.330\hsize]{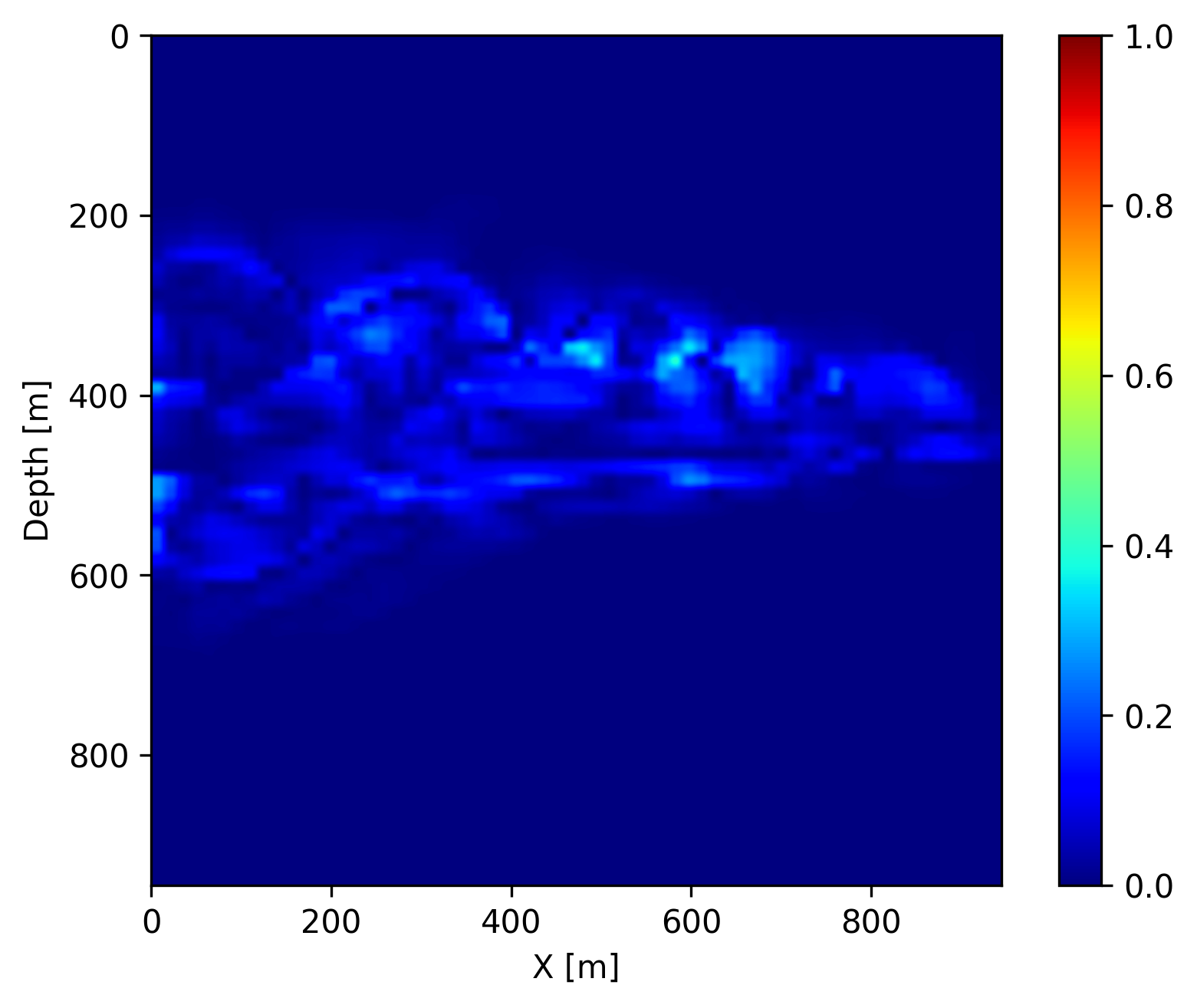}}
\subfloat[\label{fig:diff-5}]{\includegraphics[width=0.330\hsize]{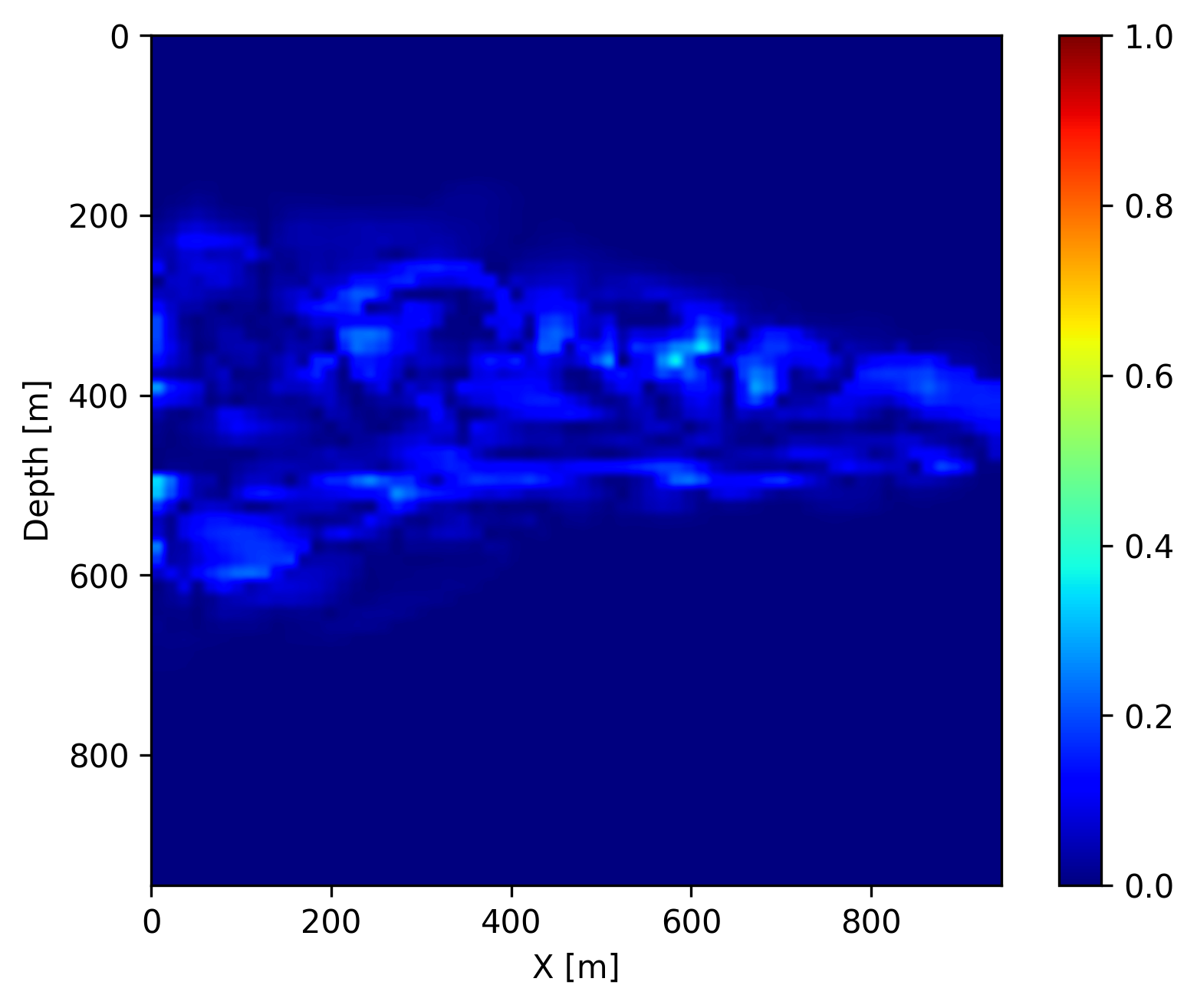}}
\subfloat[\label{fig:diff-6}]{\includegraphics[width=0.330\hsize]{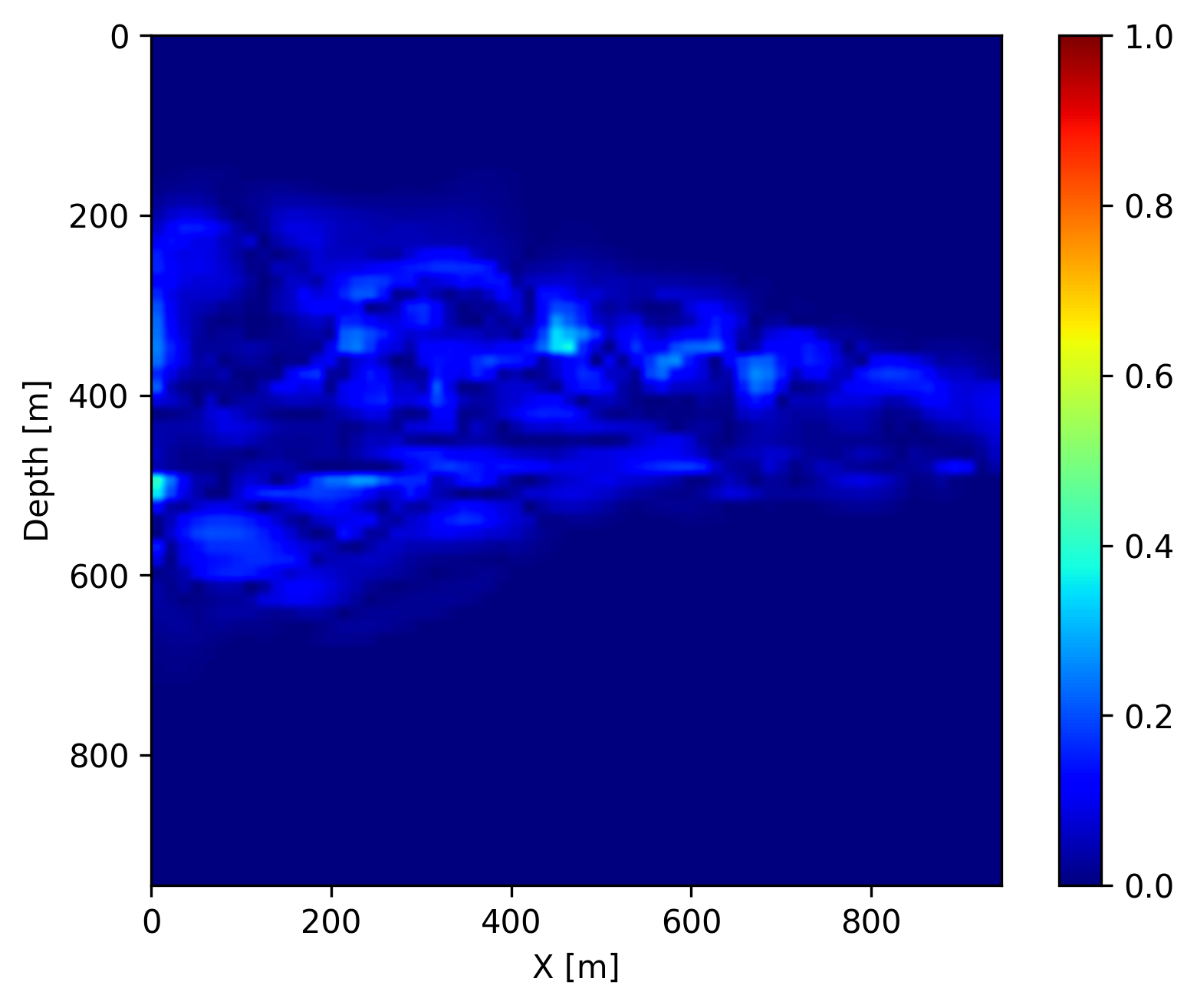}}
\caption{(a)(b)(c) CO\textsubscript{2} concentration forecast at $440$,
$560$, and $680$ days after injection. (d)(e)(f) Ground truth snapshots
of CO\textsubscript{2} concentration from numerical simulation on the
ground truth permeability model. (g)(h)(i) Difference plotted in the
same scale.}\label{fig:forecast}
\end{figure}

Finally, once we reach the $400$ days, we do not have access to seismic
data anymore (in the future). However, we can use the trained FNO to
forecast the CO\textsubscript{2} concentration in the relatively near
future assuming that the fluid dynamic will not drastically change. We
show on Figure~\ref{fig:forecast} the forecast at day $440$, $560$ and
$680$ by juxtaposing them with the ground truth obtained through
numerical simulation and the differences. We observe that the forecast
is relatively accurate and catches the global behavior of the plume even
though no seismic monitoring data is available. This result confirms
that the pre-trained FNO in combination with the permeability estimate
from the time-lapse seismic can provide a forecasting framework for
seismic monitoring of geological carbon storage.

The learned coupled inversion framework is implemented in Julia, where
we use \href{https://github.com/FluxML/Flux.jl}{Flux.jl} for AD. The
scripts to reproduce the experiments are available on the SLIM GitHub
page \url{https://github.com/slimgroup/FNO4CO2}.

\vspace*{-0.3cm}

\section{Discussion and conclusion}\label{discussion-and-conclusion}

\vspace*{-0.2cm}

Coupled inversion for carbon sequestration monitoring is computationally
challenging as it needs to iteratively solve fluid-flow and wave
equations, and differentiate through the solvers. We overcome this
problem by replacing the fluid-flow solver by a pre-trained Fourier
neural operator, which reduces the computational cost of fluid-flow
simulations and differentiation. We demonstrated that the learned
coupled inversion framework can yield reasonable estimates of the
permeability of the reservoir. This estimated permeability can then be
used for not only generating the CO\textsubscript{2} concentration
snapshots at the current vintages, but also forecasting the growth of
the CO\textsubscript{2} plume in the future. This can potentially enable
uncertainty quantification for potential plume behaviors in the future
for risk management. While these initial results on learned coupled
inversion are encouraging, more realistic physics phenomena can be
considered in future work to numerically model the fluid-flow and wave
physics more accurately. More robust inversion methods with
regularization and constraints may also lead to better estimation of the
permeability and CO\textsubscript{2} concentration. Future work will
also involve exploration of the generalization capability of the Fourier
neural operator and development of a large-scale $3$D continuous
monitoring framework that potentially updates the permeability according
to the latest acquired seismic data from the field.

\vspace*{-0.3cm}

\section{Acknowledgement}\label{acknowledgement}

\vspace*{-0.2cm}

This research was carried out with the support of Georgia Research
Alliance and partners of the ML4Seismic Center. The authors thank
Philipp A. Witte at Microsoft for the constructive discussion.

\bibliography{paper}

\end{document}